# A pre-metric formulation of Fresnel's wave and ray surfaces

D. H. Delphenich ([†])
Spring Valley, OH, USA

**Abstract**: The conventional formulation of the Fresnel wave and ray surfaces typically involves the implicit use of Euclidian metrics in order to justify treating vectors and covectors as indistinguishable. This tends to disguise the fact that one of the surfaces lives in the cotangent spaces to the spatial (or space-time) manifold, while the other lives in the tangent spaces. Moreover, no mention is made of how to get from tangent spaces to cotangent spaces in the absence of a metric. The following analysis shows how to resolve those issues within the framework of pre-metric electromagnetism. The way that electromagnetic waves imply spatial frames, coframes, and metrics will be explained.

**Contents**

————

**Introduction.** – The concept of pre-metric electromagnetism is more than a specialized technique that relates to only the theory of electromagnetism, it is a major paradigm shift in terms of what theoretical physics considers to be the most-fundamental structures on the spacetime manifold. Currently, in the wake of the stellar success of Einstein's general theory of relativity, most theoretical physicists seem to believe that the most-fundamental structure on the spacetime manifold is the Lorentzian metric that is presumed to exist in each of its tangent spaces. That metric tensor then defines a canonical connection on the bundle of Lorentzian frames in the form of the Levi-Civita connection, whose associated Riemannian curvature 2-form is then (indirectly) coupled to the distribution of energy, momentum, and stress in space by way of Einstein's equations of gravitation. The paths of both massive and massless matter are then determined from the geodesic equations that the connection defines.

However, that basis for the existence of all physical phenomena seems to have mostly advanced the cause of astrophysics and cosmology more than the more-terrestrial physics that is largely electromagnetic in nature, until one probes the internal structure of atoms, nuclei, and nucleons, when it is traditional to introduce strong and weak interactions, which only exist at that scale of

———————————————
([†]) E-mail: feedback@neo-classical-physics.info, Website: neo-classical-physics.info.

phenomena. So far, it is quantum field theory, such as quantum electrodynamics, electroweak theory, and quantum chromodynamics, that has been most successful in dealing with the realm of quantum phenomena (i.e., at the atomic to subatomic scale). Gravitation, by contrast, has always been regarded as largely irrelevant at that scale. Thus, an ongoing problem in theoretical physics has been to reconcile the approach that is taken in order to model quantum phenomena with the approach that one takes in order to model astrophysical and cosmological ones. That is generally the goal of the so-called "quantum gravity." Typically, the Holy Grail that one seeks would be a "Theory of Everything," in the form of a unified (most-likely gauge) field theory in which all four fundamental interactions would be merely different facets of the same multifaceted gemstone.

However, from the standpoint of pre-metric electromagnetism, the spacetime metric is only a corollary to something more fundamental in the structure of spacetime, namely, its electromagnetic constitutive laws. The corollary then comes about when one looks at the dispersion law for the propagation of electromagnetic waves. As it turns out, the Lorentzian metric, which is a generalization of the dispersion law for the classical electromagnetic vacuum, is defined by the most elementary constitutive law that one can realistically imagine. Thus, the Lorentzian metric is merely a generalization of a specialization that ignored a vast number of alternative constitutive laws and dispersion laws that are still quite well established in experimental physics and relate to such well-developed fields as optics in its various guises and the propagation of electromagnetic waves in plasmas, such as the ionosphere. Ultimately, one might be looking towards a possible expansion in scope of the classical electromagnetic vacuum to the quantum electromagnetic vacuum, in which vacuum polarization seems to play a recurring fundamental role.

A question then arises when one attempts to start with the electromagnetic structure of spacetime in regard to what kind of geometry one would be doing in the absence of metrics, along with the question of how the metrics first come about. The author has previously discussed the role of line geometry as the most-natural geometry for electromagnetism in general [1]. In this article, the topic of pre-metric geometry will be specialized to a largely optical class of constitutive laws in order to show that the very existence of "wave-like solutions" to the pre-metric Maxwell equations will imply a canonical 3-coframe in the spatial cotangent spaces that basically amounts to the usual **k**, **e**, **h** frame in the more conventional vector-analytic approach to electromagnetism (such as one finds in Sommerfeld [2]) and a reciprocal 3-frame in the spatial tangent spaces that corresponds to the usual **s**, **d**, **b** frame. That will then define a linear isomorphism of cotangent spaces with tangent spaces in the process. The reciprocal frame and coframe will then be used to define the Fresnel wave surface in the cotangent spaces as a dispersion law for the frequency-wave number 1-form *k* and its corresponding ray surface in the tangent spaces as a constraint surface of a more kinematical nature that relates to the ray vectors, which are tangent to the light rays.

The novelty of the present approach is that it frees the theoretician from the limitations of vector analysis, which is the usual way of discussing the electrodynamics of continuous media (cf., e.g., [3]), and which assumes the existence of numerous "accidental" linear isomorphisms that are peculiar to three dimensions. In particular, one typically does not distinguish between vectors and covectors, and only distinguishes between vectors and bivectors by referring to them as "polar" and "axial" vectors, respectively. Similarly, bivectors and 2-forms would also become indistinguishable from vectors and covectors. As a result of that resolution of the fine details, the problem of defining linear isomorphisms between the four different three-dimensional vector spaces becomes more subtle. It will be shown that two of the isomorphisms can be obtained from the Poincaré duality that is defined by a volume element on spacetime (which is then assumed to be orientable), while the other one, namely, the isomorphism of tangent and cotangent spaces, can be defined by the canonical frames and coframes that any electromagnetic wave defines, regardless of constitutive law.

Thus, this discussion will begin with a brief review of the relevant topics in pre-metric electromagnetism in section **1**, which is followed by a discussion of electromagnetic waves in section **2**. That discussion is specialized to the case of the Fresnel wave and ray surfaces in section **3**. The main discussion of canonical frames and metrics that are defined by electromagnetic waves is then found in section **4**, and some concluding remarks are made in section **5**. One of the more lengthy calculations was moved to an Appendix in order to not distract from the general line of reasoning.

**1. Pre-metric electromagnetism.** – In most modern treatments of the formulation of Maxwell's equations in the language of exterior differential forms [**4**], one is implicitly dealing with the simplest of all possible electromagnetic media, namely, the classical electromagnetic vacuum. To some extent, that is a natural consequence of the fact that the earliest discussions of special relativity were concerned with the way that electromagnetic waves would propagate in media that were in a state of relative motion, but the media in question were generally as simple as possible. As a result, when the purely-mathematical aspects of the special relativistic formulation of Maxwell's equations were successively generalized, they were usually generalizations of a single example that tended to overlook all of the other examples of electromagnetic media in nature and the laboratory.

Although in terms of the practical applications of Maxwell's equations, the introduction of electromagnetic media that were more involved than the classical vacuum was commonplace, such as the entire field of optics and the study of the propagation of electromagnetic waves in plasmas, such as the Earth's ionosphere, it was perhaps the quantum theory of fields that first started pointing to the possible breakdown of the simplicity that relativity theory was based in. That is because a recurring theme in quantum field theory, and quantum electrodynamics in particular, was the fundamental role played by "vacuum polarization." That takes the form of the formation of "virtual" particle-antiparticle pairs at sufficiently-high field strengths, and typically they are introduced into the perturbation expansions (in particular, the "loop" expansions) of the scattering amplitudes for particle interactions in the form of loops on internal branches in the Feynman diagrams. Thus, the loop expansions are giving the theoretician possibly the most promising Ansatz for generalizing the first principles of electromagnetism, namely, generalizing the structure of the classical vacuum to the quantum vacuum.

*a. Electromagnetic constitutive laws*. – A first step in that direction is to generalize the formulation of Maxwell's equations from the form that they take for the classical vacuum to their more general form for continuous electromagnetic media. Mostly, that generalization revolves around the electromagnetic constitutive properties of the medium. In the conventional formulation of the electrodynamics of continuous media [**3**], that takes the form of an operator that takes the basic electric and magnetic field strengths, which take the form of spatial vector fields **E** and **H**, to the electric and magnetic excitation vector fields **D** and **B**, which include the possibility that the **E** and **H** fields have induced the formation of electric and magnetic dipoles in the medium, which is one of the two ways that the word "polarization" gets used in the electrodynamics of continuous media ([1]).

Just to assume that the relationship between the fields **E** and **H** and the excitations **D** and **B** that they produce is algebraic, and not an integral operator, is already a first simplification since it is possible that the excitation that results from the **E** and **H** fields at one point depends upon the state of excitation at the neighboring points in both space and time. That sort of non-locality is one of the two inequivalent uses of the word "dispersion" in the electrodynamics of continuous media. The other usage, which we shall be dealing with here, relates to a constraint hypersurface that the frequency-wave number 1-form must lie on, and which eventually accounts for the light cones of special relativity that get generalized to the Lorentzian metric of general relativity, which in turn implies the presence of gravitation in space-time.

Having made the first reduction in scope by assuming that the constitutive laws of the medium takes the form of an invertible algebraic operator, and not an integral one (so the medium is "non-dispersive"), one can then express those laws in the form:

(1.1) $$\mathbf{D} = \varepsilon(\mathbf{E}, \mathbf{H}), \qquad \mathbf{B} = \mu(\mathbf{E}, \mathbf{H}),$$

which can then be solved for:

(1.2) $$\mathbf{E} = \varepsilon^{-1}(\mathbf{D}, \mathbf{B}), \qquad \mathbf{H} = \mu^{-1}(\mathbf{D}, \mathbf{B}).$$

A second reduction would be to assume that the invertible algebraic operator is also an invertible *linear* operator. Therefore, if one chooses some suitable (possibly local) frame field $\{\mathbf{e}_i, i = 1, 2, 3\}$ on space and expresses the vector fields in terms of their components with respect to that frame field:

(1.3) $$\mathbf{E} = E^i(t, x^j)\mathbf{e}_i, \quad \mathbf{D} = D^i(t, x^j)\mathbf{e}_i, \quad \mathbf{H} = H^i(t, x^j)\mathbf{e}_i, \quad \mathbf{B} = B^i(t, x^j)\mathbf{e}_i$$

then the linear constitutive law can be expressed in the matrix form:

(1.4) $$D^i = \varepsilon^i_j(t, x^k) E^j + \alpha^i_j(t, x^k) H^j, \qquad B^i = \bar{\alpha}^i_j(t, x^k) E^j + \mu^i_j(t, x^k) H^j.$$

The matrix $\varepsilon^i_j(t, x^k)$ represents the *dielectric strength* tensor of the medium. The matrices $\alpha^i_j(t, x^k)$ and $\bar{\alpha}^i_j(t, x^k)$ account for possible cross-couplings of electric and magnetic fields, such as ones that arise by a particular choice of frame field in that same way that "fictitious" accelerations can arise in rotational mechanics, Faraday rotations, and optical activity in the media ([1]). The matrix $\mu^i_j(t, x^k)$ represents the *magnetic permeability* of the medium. In the nonlinear case, which is of course more relevant to the high field strengths of quantum electrodynamics, the matrices would also depend upon the components of **E** and **B**, as well. Nonlinear electromagnetism is also unavoidable when one goes from linear optics to nonlinear optics.

---

[1] The other, largely unrelated, usage of the work "polarization" refers to the fact that the **E** and **H** vectors for an electromagnetic wave define an orthonormal 2-frame whose angular position in the plane that they span can vary as the wave propagates. We will not be dealing with that aspect of electromagnetic waves, except insofar as we shall discuss the nature of the canonical frames that electromagnetic waves define.

In order to get to the typical (i.e., linear) optical media, which will be the main focus of the following discussion, one further specializes the medium to one in which there are no cross-couplings, and the magnetic polarization of the medium is negligible. That has the effect of reducing the constitutive laws to the form:

(1.5) $$D^i = \varepsilon^i_j(t, x^k) E^j, \qquad B^i = \mu_0 \delta^i_j H^j = \mu_0 H^i.$$

The matrix $\varepsilon^i_j(t, x^k)$ is generally assumed to be symmetric, so it will also be diagonalizable. That is, there will be a special frame called a *principal frame* for $\varepsilon^i_j(t, x^k)$ in which it takes the form:

(1.6) $$\varepsilon^i_j(t, x^k) = \text{diag}\,[\varepsilon_1, \varepsilon_2, \varepsilon_3],$$

in which the diagonal elements $\varepsilon_i$, $i = 1, 2, 3$ are also functions of $t$ and $x^k$, and they are called the *principal dielectric strengths* of the medium at each point and time. Further reductions in generality are possible, such as assuming the components of the matrix $\varepsilon^i_j(t, x^k)$ are time-invariant or constant in space or both. The medium would then be called *electrically homogeneous*. Clearly, such a reduction would have to depend upon the choice of frame field that is used to define the components, which would otherwise seem to contradict the spirit of relativity, but we have not yet reached the point where we can properly discuss the scope of special relativity. Another reduction in generality is to assume that the diagonal elements of $\varepsilon^i_j(t, x^k)$ in its principal frame are all equal to each other, and in that case, one calls the medium *electrically isotropic*. An intermediate step is to assume that two of them are equal, while the third one is different, and such a medium is referred to as *uniaxial*, as opposed to the general case, which is *biaxial*.

Thus, one is also assuming that the magnetic properties of the medium are those of a non-dispersive, linear, homogeneous, isotropic magnetic medium. If one assumes that the medium has same properties as an electric medium then the constitutive law will take the simplest of all possible forms:

(1.7) $$\mathbf{D} = \varepsilon_0\,\mathbf{E}, \qquad \mathbf{B} = \mu_0\,\mathbf{H}.$$

---

([1]) Since we shall not be concerned with such phenomena in what follows, the curious can confer some of the standard literature on optics, such as [2] and [5].

That is the type of constitutive law that is the basis for special relativity and is referred to as the *classical electromagnetic vacuum*. Generally, the only way that it enters into the Maxwell equations (at least in their vector form) is by way of the speed of propagation of electromagnetic waves in that medium, i.e., the speed of light *in vacuo*:

$$(1.8) \qquad c = \frac{1}{\sqrt{\varepsilon_0 \mu_0}}.$$

It is interesting that although the constancy of $\varepsilon_0$ and $\mu_0$ are unavoidably frame-dependent, nonetheless, one of the foundations of special relativity is that $c$ itself should be independent of any choice of Lorentzian frame.

*b. Maxwell's equations in space-time form.* – Ordinarily, when one encounters the relativistic formulation of Maxwell's equations in terms of exterior differential forms, no mention is made of the fact that they are specialized to the electromagnetic constitutive law of the classical vacuum. Typically, that law enters into the equations tacitly, since it contributes to only the constant $c$, which is often set equal to 1 on the assumption that it is not important to be consciously aware of the speed with which electromagnetic waves propagate in the chosen medium if it is always merely the same constant everywhere and for all time. Similarly, although one often treats air as essentially the same as a vacuum, if one considers the optics of the atmosphere, one finds that even that is an oversimplification that overlooks such phenomena as clouds, rainbows, mirages, refraction at the horizon, and scattering. More precisely, the value of $c$ for air varies with time, position, and mostly with temperature, which is itself a function of time and position.

Those equations generally take the form:

$$(1.9) \qquad d_\wedge F = 0, \qquad \delta F = -4\pi J, \qquad \delta J = 0.$$

The 2-form $F$ on the spacetime manifold $M$ represents the combined electric and magnetic field strengths, and the 1-form $J$ represents the combined electric charge density and electric current that serves as the source of the field $F$. The codifferential operator $\delta$ is essentially "dual" to the exterior derivative operator $d_\wedge$ and represents a generalization of the divergence operator that acts upon vector fields to something that acts upon exterior differential forms. Thus, it will take $k$-forms to $(k-1)$-forms.

The definition of $\delta$ makes it adjoint to $d_\wedge$ with respect to the *Hodge duality* operator $* : \Lambda^k M \to \Lambda^{n-k} M$, namely:

$$(1.10) \qquad \delta = *^{-1} \cdot d_\wedge \cdot *.$$

Since one then has:

$$(1.11) \qquad \delta^2 = 0,$$

which is a consequence of the property of $d_\wedge$ that makes its own square equal to zero identically, the last equation in (1.9) represents a compatibility (or integrability) condition for $J$ to be a legitimate source 1-form.

Now, it was observed in the early days of the relativistic formulation of Maxwell's equations that the only place in which the Lorentzian metric $g$ on spacetime enters into Maxwell's equations is by way of the operator *. More precisely, as long as one assumes that the spacetime manifold $M$ is orientable and one has chosen a volume element $V \in \Lambda^4 M$, one can define it to be the composition of two isomorphisms. One first maps $k$-forms to $k$-vector fields (i.e., the exterior algebra that is defined by vector fields on $M$) using the inverse of the isomorphism $\iota_g : \Lambda_k M \to \Lambda^k M$ that one gets from the metric $g$. Namely, one first defines the isomorphism $\iota_g : T_x(M) \to T_x^*(M)$ that takes any tangent vector $\mathbf{v}$ at $x \in M$ to the 1-form $\iota_g(\mathbf{v})$ that gives:

$$(1.12) \qquad \iota_g(\mathbf{v})(\mathbf{w}) = g(\mathbf{v}, \mathbf{w}) = g_{\mu\nu} v^\mu w^\nu$$

whenever it is evaluated on any tangent vector $\mathbf{w}$ at $x$.

Typically, in the component formulation of tensor calculus, that isomorphism $\iota_g$ is referred to as "lowering an index," since it will take the components of $\mathbf{v}$, namely, $v^\mu$, to the components of $\iota_g(\mathbf{v})$:

$$(1.13) \qquad v_\mu = g_{\mu\nu} v^\nu.$$

In order to extend that isomorphism to all $k$-vector fields, one can apply it to each vector in an exterior product of $k$ vectors $\mathbf{v}_1 \wedge \ldots \wedge \mathbf{v}_k$ to get the $k$-form:

$$(1.14) \qquad \iota_g(\mathbf{v}_1 \wedge \ldots \wedge \mathbf{v}_k) = \iota_g(\mathbf{v}_1) \wedge \ldots \wedge \iota_g(\mathbf{v}_k).$$

Since any $k$-vector is a finite linear combination of "decomposable" $k$-vectors with the same form as $\mathbf{v}_1 \wedge \ldots \wedge \mathbf{v}_k$, one then extends the isomorphism $\iota_g$ "by linearity." That is, one defines its effect on a linear combination of decomposable $k$-vectors to be that same linear combination of decomposable $k$-forms. The effect of that map on components is to "lower all indices" of the $k$-vector field. That is, if the $k$-vector field takes the component form ([1]):

$$(1.15) \qquad \mathbf{B} = B^{\mu_1 \cdots \mu_k} \partial_{\mu_1} \wedge \cdots \wedge \partial_{\mu_k}$$

then the components of the $k$-form $\iota_g(\mathbf{B})$ will be:

---

([1]) From now on, components of tensor fields, such as multivector fields and differential forms, will generally be defined with respect to natural frame and coframe fields that are defined by local coordinate charts. That is, if $(U, x^\mu)$ is a local coordinate chart about a point in the spacetime manifold $M$ then the natural frame field that it defines is the set of tangent vector fields $\{\partial_\mu = \partial/\partial x^\mu, \mu = 0, 1, 2, 3\}$, and the natural coframe field is the set of 1-forms $\{dx^\mu, \mu = 0, 1, 2, 3\}$.

(1.16) $$B_{\mu_1 \cdots \mu_k} = g_{\mu_1 \nu_1} \cdots g_{\mu_k \nu_k} B^{\nu_1 \cdots \nu_k}.$$

The inverse isomorphism $\iota_g^{-1}: \Lambda^k M \to \Lambda_k M$ basically amounts to "raising all indices" of a $k$-form. In particular, for the case in question of the 2-form $F$, one will get the bivector field $\iota_g^{-1}(F)$ whose components are usually described by:

(1.17) $$F^{\mu\nu} = g^{\mu\kappa} g^{\nu\lambda} F_{\kappa\lambda}.$$

More precisely, since the $F_{\mu\nu}$ is antisymmetric in its indices, as is $F^{\mu\nu}$, one must antisymmetrize the last equation:

(1.18) $$F^{\mu\nu} = \tfrac{1}{2}(g^{\mu\kappa} g^{\nu\lambda} - g^{\mu\lambda} g^{\nu\kappa}) F_{\kappa\lambda}.$$

One can then think of the linear isomorphism $\iota_g^{-1}$, as it acts upon 2-forms, as having the components:

(1.19) $$g^{\mu\nu\kappa\lambda} = \tfrac{1}{2}(g^{\mu\kappa} g^{\nu\lambda} - g^{\mu\lambda} g^{\nu\kappa}).$$

The second isomorphism that was alluded to above in the decomposition of * is the Poincaré duality isomorphism $\# : \Lambda_k M \to \Lambda^{4-k} M$, which takes the $k$-vector field **B** to the $(4-k)$-form that amounts to the interior product of the volume element $V$ with the $k$-vector field **B**:

(1.20) $$\#\mathbf{B} = i_\mathbf{B} V.$$

If the volume element is expressed in component form as:

(1.21) $$V = dx^0 \wedge dx^1 \wedge dx^2 \wedge dx^3 = \frac{1}{4!} \varepsilon_{\mu_0 \mu_1 \mu_2 \mu_3} dx^{\mu_0} \wedge dx^{\mu_1} \wedge dx^{\mu_2} \wedge dx^{\mu_3}$$

then the components of the 2-form $\#\mathbf{B}$, when **B** is a bivector field, will take the form:

(1.22) $$(\#\mathbf{B})_{\mu\nu} = \tfrac{1}{2} \varepsilon_{\mu\nu\kappa\lambda} B^{\kappa\lambda}.$$

We can then decompose * into the composition of linear isomorphisms:

(1.23) $$* = \# \cdot \iota_g^{-1}.$$

This is the point of departure for "pre-metric" electromagnetism [6]. Basically, one replaces the isomorphism $\iota_g^{-1}$ with an electromagnetic constitutive law $C : \Lambda^2 M \to \Lambda_2 M$, which will associate the 2-form $F$ with a bivector field $\mathfrak{H}$ that represents the electric and magnetic excitations of the medium that are induced by the electric and magnetic fields, i.e., the formation of electric and magnetic dipoles in response to the presence of those fields:

(1.24) $$\mathfrak{H} = C(F), \quad \text{i.e.,} \quad \mathfrak{H}^{\mu\nu} = \tfrac{1}{2} C^{\mu\nu\kappa\lambda} F_{\kappa\lambda}.$$

More precisely, $C$ must be a *linear* electromagnetic constitutive law, since nonlinear electromagnetism would have to be recast in terms of 2-forms on the *total space* $\Lambda^2 M$, not 2-forms on $M$.

One sees that there is no longer any need to introduce a Lorentzian metric at this point. To relate the isomorphism just defined to the one that was defined by a Lorentzian metric above, one would simply replace $C$ with $\iota_g^{-1}$, i.e., one would set:

(1.25) $$C^{\mu\nu\kappa\lambda} = g^{\mu\nu\kappa\lambda}.$$

In order to recast Maxwell's equations in their pre-metric form, we need to introduce one further operator that generalizes the divergence of vector fields, but acts upon $k$-vector fields, instead of $k$-forms, as the codifferential does. Namely, for each $k$, the generalized divergence operator is a linear map div : $\Lambda_k \to \Lambda_{k-1}$ that is defined to the adjoint of $d_\wedge$ with respect to # :

(1.26) $$\text{div} = \#^{-1} \cdot d_\wedge \cdot \#.$$

One can then see that this operator actually agrees with the usual divergence of vector fields, namely, if the vector field has the local form $\mathbf{X} = X^\mu \partial_\mu$ then one will have:

(1.27) $$\text{div } \mathbf{X} = \frac{\partial X^\mu}{\partial x^\mu}.$$

Although the operator div is linear on $k$-vector fields, just as $d_\wedge$ is linear on $k$-forms, and one has:

(1.28) $$\text{div} \cdot \text{div} = \#^{-1} \cdot d_\wedge \cdot \# \cdot \#^{-1} \cdot d_\wedge \cdot \# = \#^{-1} \cdot (d_\wedge)^2 \cdot \# = 0,$$

nonetheless, the divergence operator does not behave the same in regard to exterior products as the exterior derivative, that is:

(1.29) $$\text{div } (\mathbf{A} \wedge \mathbf{B}) \neq \text{div } \mathbf{A} \wedge \mathbf{B} + (-1)^k \mathbf{A} \wedge \text{div } \mathbf{B},$$

in general.

We can now recast Maxwell's equations in a form that appeals to the electromagnetic constitutive law $C$ in place of the Lorentzian metric:

(1.30)         $d_\wedge F = 0$,    div $\mathfrak{H} = 4\pi \mathbf{J}$,    div $\mathbf{J} = 0$,    $\mathfrak{H} = C(F)$.

The source field **J** is now a vector field, which is more consistent with the way that physicists typically regard currents. In particular, "current algebras" will naturally be sub-algebras of the Lie algebra $\mathfrak{X}(M)$ of vector fields on *M* then. Indeed, it is easy to see that if $\mathbf{J}_1$ and $\mathbf{J}_2$ are vector fields with zero divergence then one will also have:

(1.31)         div $[\mathbf{J}_1, \mathbf{J}_2] = 0$.

Thus, the (infinite-dimensional) vector subspace of $\mathfrak{X}(M)$ that is defined by zero-divergence vector fields is closed under Lie bracket. Indeed that should not be surprising, since such vector fields are infinitesimal generators of one-parameter families of volume-preserving diffeomorphisms of *M*, and such diffeomorphisms form a subgroup of the group of all diffeomorphisms of *M*.

As we shall see in the next section, the Lorentzian metric will eventually reappear in the context of pre-metric electromagnetism as a degenerate case of a dispersion law for the propagation of electromagnetic waves.

*c. Time+space decomposition*. – Virtually all textbooks and articles that deal with the electrodynamics of continuous media discuss the law of electromagnetism in the traditional formalism of vector calculus. In order to show that the more unconventional formulation in the subsection above is consistent with those discussions, it is necessary to introduce the spatial 1-forms *E*, *H* and the spatial vector fields **D**, **B**. That, in turn, requires that the four-dimensional spacetime (or at least its tangent and cotangent bundles) must have some physically-meaningful splitting into a time and a space part.

Although we could generalize the natural splitting of local coordinates $\{x^\mu, \mu = 0, 1, 2, 3\}$ with $x^0 = ct$ into $\{x^0, x^i\}$, since we are really more concerned about infinitesimal things, i.e., tangent and cotangent objects, it is mostly sufficient for us to have direct sum splittings of $T(M)$ and $T^*(M)$:

(1.32)         $T(M) = L(\mathbf{t}) \oplus \Sigma(M)$,    $T^*(M) = L(\tau) \oplus \Sigma^*(M)$,

in which **t** is an everywhere-non-zero vector field, so $L(\mathbf{t})$ is the line field on *M* that it generates, and $\tau$ is an everywhere-non-zero 1-form, so $L(\tau)$ is the line field in $T^*(M)$ that it generates. As long as $\tau$ and **t** are transverse, in the sense that:

(1.33)         $\tau(\mathbf{t}) \neq 0$,

one can then define the spatial sub-bundles $\Sigma(M)$ and $\Sigma^*(M)$ by saying that their fibers are the vector subspaces that are annihilated by $\tau$ in the former case and annihilated by **t** in the latter. That is, if $\mathbf{X} \in \Sigma(M)$ and $\alpha \in \Sigma^*(M)$ then:

(1.34) $$\tau(\mathbf{X}) = 0 \quad \text{and} \quad \alpha(\mathbf{t}) = 0.$$

The pair $(\mathbf{t}, \tau)$ is sometimes ([1]) referred to as a choice of *observer*, since the vector field $\mathbf{t}$ represents the timeline for that observer, and also defines the rest space for motion relative to the observer. Dually, $\tau$ defines the complementary spatial subspaces of the tangent spaces along the timeline of that observer.

The splittings of $T(M)$ and $T^*(M)$ induce corresponding splittings in all of the tensor products of those vector bundles. In particular, the bundles of 2-forms and bivector fields split into:

(1.35) $$\Lambda^2 = \tau^2 \oplus \Sigma^2 \quad \text{and} \quad \Lambda_2 = \mathbf{t}_2 \oplus \Sigma_2.$$

The bundle $\tau^2$ has three-dimensional fibers. If the natural coframe field $\{dx^\mu, \mu = 0, 1, 2, 3\}$ is adapted to the splitting of $\Lambda^2$, so:

(1.36) $$\tau = dx^0 = c\, dt,$$

and $\{dx^i, i = 1, 2, 3\}$ span the fibers of $\Sigma(M)$, then the 2-forms in $\tau^2$ will take the local form:

(1.37) $$\alpha = c\, dt \wedge \beta, \qquad \beta = \beta_i(t, x^j)\, dx^i,$$

in which $\beta$ is a 1-form in $\Sigma(M)$, although one should note that its components are functions of space and time, and not just space.

The bundle $\Sigma^2$ of spatial 2-forms also has three-dimensional fibers. Indeed, it can be shown to coincide with the bundle of 2-forms over S:

$$\Sigma^2 = \Lambda^2(\Sigma^*).$$

Thus, the spatial 2-forms will take the local form:

(1.38) $$\gamma = \tfrac{1}{2} \gamma_{ij}(t, x^k)\, dx^i \wedge dx^j.$$

Once again, it should be noted that the components are functions of space and time.

One has a dual situation going on in regard to the splitting of $\Lambda_2 M$, namely, both $\mathbf{t}_2$ and $\Sigma_2$ have three-dimensional fibers, bivectors in $\mathbf{t}_2$ take the form:

(1.39) $$\mathbf{B} = \frac{1}{c} \partial_t \wedge \mathbf{X}, \qquad \mathbf{X} = X^i(t, x^j)\, \partial_i,$$

---

([1]) The author has written several articles [7] on the subject of how time+space splittings of the tangent bundle and cotangent bundles relate to fundamental physics and other mathematics, such as Ehresmann connections.

and $\Sigma_2 = \Lambda_2(\Sigma)$, so its bivectors take the local form:

$$(1.40) \qquad \mathbf{B} = \tfrac{1}{2} B^{ij}(t, x^k) \partial_i \wedge \partial_j.$$

Consequently, the electromagnetic field strength 2-form can be expressed in time + space form as:

$$(1.41) \qquad F = c\, dt \wedge E - B,$$

in which the spatial 1-form $E$ and the spatial 2-form $B$ have the local forms:

$$(1.42) \qquad E = E_i(t,x)\, dx^i, \qquad B = \tfrac{1}{2} B_{ij}(t,x)\, dx^i \wedge dx^j.$$

Note that we are combining a field strength (viz., $E$) with the spatial Poincaré dual $B = \#_s \mathbf{B}$ of an excitation vector field $\mathbf{B}$. That is, since the time-space splitting of $T^*M$ induces a corresponding splitting of $\Lambda^4(M)$, the volume element $V$ can also be expressed in the form:

$$(1.43) \qquad V = \tau \wedge V_s, \qquad \tau = c\, dt, \qquad V_s = dx^1 \wedge dx^2 \wedge dx^3 = \frac{1}{3!} \varepsilon_{i_1 i_2 i_3}\, dx^{i_1} \wedge dx^{i_2} \wedge dx^{i_3}.$$

As a result, one can define Poincaré duality for the spatial multi-vectors and $k$-forms by way of the linear isomorphisms $\#_s : \Sigma_k \to \Sigma^{3-k}$ that take any spatial $k$-vector field $\mathbf{B}$ to its dual spatial $(3-k)$-form:

$$(1.44) \qquad \#_s \mathbf{B} = i_\mathbf{B}\, V_s,$$

so when $\mathbf{B}$ is a spatial vector field, its spatial dual will be a spatial 2-form whose components take the form:

$$(1.45) \qquad B_{ij} = (\#_s \mathbf{B})_{ij} = \varepsilon_{ijk} B^k.$$

When one takes the exterior derivative of $F$, one will also get a time+space decomposition of the result:

$$d_\wedge F = -c\, dt \wedge d_\wedge E - d_\wedge B = -c\, dt \wedge d_\wedge E - d_\wedge (\#_s \mathbf{B}).$$

Furthermore:

$$d_\wedge E = \partial_t E \wedge dt + d_{\wedge s} E = -dt \wedge \partial_t E + \tfrac{1}{2}(\partial_i E_j - \partial_j E_i)\, dx^i \wedge dx^j$$

and

$$d_\wedge (\#_s \mathbf{B}) = \partial_t (\#_s \mathbf{B}) \wedge dt + d_{\wedge s}\, \#_s \mathbf{B} = \partial_t (\#_s \mathbf{B}) \wedge dt + \#_s \operatorname{div}_s \mathbf{B},$$

in which we have defined the spatial divergence operator in the obvious way:

(1.46) $$\text{div}_s = \#_s^{-1} \cdot d_{\wedge s} \cdot \#_s .$$

When we combine all of the specialized results above, we can rewrite $d_\wedge F$ in the form:

(1.47) $$d_\wedge F = - dt \wedge [c \, d_{\wedge s} E + \partial_t (\#_s \mathbf{B})] - \#_s \, \text{div}_s \, \mathbf{B} ,$$

so its vanishing would be equivalent to the pair of equations:

(1.48) $$d_{\wedge s} E = - \frac{1}{c} \frac{\partial}{\partial t} (\#_s \mathbf{B}) , \qquad \text{div}_s \, \mathbf{B} = 0 ,$$

which have the form of two of the usual Maxwell equations:

(1.49) $$\nabla \times \mathbf{E} = - \frac{1}{c} \frac{\partial \mathbf{B}}{\partial t} , \qquad \nabla \cdot \mathbf{B} = 0 .$$

In order to get the other two, we begin by decomposing the bivector field $\mathfrak{H}$ into time+space form:

(1.50) $$\mathfrak{H} = \frac{1}{c} \partial_t \wedge \mathbf{D} + \mathbf{H} = \frac{1}{c} \partial_t \wedge \mathbf{D} + \#_s^{-1} H .$$

Taking the divergence of that expression requires slightly more special handling. Since the calculations would be a distraction at this point, we defer them to an Appendix and simply give the final result.

(1.51) $$\text{div} \, \mathfrak{H} = - \frac{1}{c} \partial_t \mathbf{D} + \#_s^{-1}(d_{\wedge s} H) + \text{div}_s \, \mathbf{D} \, \frac{1}{c} \partial_t .$$

If we represent the source current in time+space form as:

(1.52) $$\mathbf{J} = \frac{\rho}{c} \partial_t + \mathbf{j} = \frac{1}{c} \rho(t,x) \partial_t + j^i(t,x) \partial_i$$

then the second Maxwell equation will become:

$$- \frac{1}{c} \partial_t \mathbf{D} + \#_s^{-1}(d_{\wedge s} H) + \text{div}_s \, \mathbf{D} \, \frac{1}{c} \partial_t = \frac{1}{c} 4\pi \rho \partial_t + 4\pi \mathbf{j} ,$$

which splits into the pair of equations:

(1.53) $$\#_s^{-1}(d_{\wedge s} H) = \frac{1}{c} \frac{\partial \mathbf{D}}{\partial t} + 4\pi \mathbf{j} , \qquad \text{div}_s \, \mathbf{D} = 4\pi \rho .$$

These have the form of the other two Maxwell equations in vector form:

(1.54) $$\nabla \times \mathbf{H} = \frac{1}{c}\frac{\partial \mathbf{D}}{\partial t} + 4\pi \mathbf{j}, \qquad \nabla \cdot \mathbf{D} = 4\pi \rho.$$

The compatibility condition for **J** takes the usual time+space form:

(1.55) $$\mathrm{div}\,\mathbf{J} = \frac{\partial \rho}{\partial t} + \mathrm{div}_s\,\mathbf{j} = 0.$$

We then summarize the results of the conversion of the Maxwell equations into time+space form:

(1.56) 
$$\boxed{\begin{aligned} d_{\wedge s} E &= -\frac{1}{c}\frac{\partial (\#_s \mathbf{B})}{\partial t}, & \mathrm{div}_s\,\mathbf{B} &= 0, \\ \#_s^{-1}(d_{\wedge s} H) &= \frac{1}{c}\frac{\partial \mathbf{D}}{\partial t} + 4\pi\mathbf{j}, & \mathrm{div}_s\,\mathbf{D} &= 4\pi\rho, & \frac{\partial \rho}{\partial t} = -\mathrm{div}_s\,\mathbf{j}. \end{aligned}}$$

One can also put $F(\mathfrak{H})$ into time+space form:

(1.57) $$F(\mathfrak{H}) = E(\mathbf{D}) - B(\mathbf{H}) = E(\mathbf{D}) - H(\mathbf{B}),$$

which can be shown to take the form of the Lagrangian density of the electromagnetic field.

**2. Electromagnetic waves.** – Although electromagnetic waves, like any other fields, naturally need to have sources that produce them, nonetheless, in order to get from the field equations for electromagnetic fields in general to the equations for electromagnetic waves in particular, one begins by considering only those points of the medium in which they are to propagate that do not represent source points. The field equations will then reduce to:

(2.1) $$d_\wedge F = 0, \qquad \mathrm{div}\,\mathfrak{H} = 0, \qquad \mathfrak{H} = C(F).$$

Although one might try to obtain a wave equation for $F$ from those equations by defining a "generalized codifferential operator," we shall not go down that route. Rather, we shall simply start from the concept of a "wave-like" solution of the field equations, which we define to be a pair of solutions to the first two equations in (2.1) that take the forms:

(2.2) $$F = e^{i\theta}\,\mathfrak{f}, \qquad \mathfrak{H} = e^{i\theta}\,\mathfrak{h},$$

in which $\theta = \theta(t, x^i)$ is a smooth function on spacetime, f is a 2-form, $\mathfrak{h}$ is a bivector field, and the last two fields satisfy the conditions that:

(2.3) $$d_\wedge f = 0, \qquad \text{div } \mathfrak{h} = 0.$$

In the usual treatment of electromagnetic waves, one introduces monochromatic plane waves, for which the components of f and $\mathfrak{h}$ would be constant in rectangular coordinates, so the last equations would be satisfied trivially.

As long as the constitutive law $C$ is linear (or at least homogeneous of degree 1), the field equations (2.1) will then take the algebraic form:

(2.4) $$k \wedge f = 0, \qquad k \wedge \#\mathfrak{h} = 0, \qquad \mathfrak{h} = C(f),$$

into which we have introduced the *frequency-wave number* 1-form:

(2.5) $$k = d\theta = \frac{\partial \theta}{\partial t} dt + \frac{\partial \theta}{\partial x^i} dx^i = \omega\, dt - k_i\, dx^i.$$

Note that the constitutive law $C$ must be consistent with the constraints (2.3). Thus, $C$ must also take closed 2-forms to bivector fields with vanishing divergence, or at least take $\#\mathfrak{h}$ to f. Usually, the way that one gets around that is to use a medium whose constitutive law is linear and homogeneous, although possibly anisotropic. We shall do that in the next section, but for now, we shall just make that comment and move on.

Just as $F$ and $\mathfrak{H}$ have time+space decompositions into the spatial fields $E$, $B$, $\mathbf{D}$, $\mathbf{H}$, the corresponding fields f and $\mathfrak{h}$ have time+space decompositions into:

(2.6) $$f = c\, dt \wedge e - b, \qquad \mathfrak{h} = \frac{1}{c} \partial_t \wedge \mathbf{d} + \mathbf{h}.$$

One can then evaluate the 2-form f on the bivector field $\mathfrak{h}$ and get a scalar function:

(2.7) $$f(\mathfrak{h}) = e(\mathbf{d}) - b(\mathbf{h}) = e_i d^i - \tfrac{1}{2} b_{ij} h^{ij},$$

which relates to the conventional Lagrangian density for the electromagnetic field.

In order to specialize the constitutive law $C$ to a time+space form, rather than use the spatial 2-form b and the spatial bivector field $\mathbf{h}$, we shall use their Poincaré spatial duals, namely, the spatial vector field $\mathbf{b}$ and the spatial 1-form h that are defined by:

(2.8) $$\mathbf{b} = \#_s^{-1} b = i_b V_s, \qquad h = \#_s \mathbf{h} = i_\mathbf{h} V_s.$$

In terms of local components, those equations take the form:

(2.9) $$b^i = \tfrac{1}{2}\varepsilon^{ijk} b_{jk}, \qquad h_i = \tfrac{1}{2}\varepsilon_{ijk} h^{jk},$$

and (2.7) will take the local form:

(2.10) $$f(\mathfrak{h}) = e_i d^i - h_i b^i.$$

We can then express the electromagnetic constitutive laws in the more conventional form:

(2.11) $$\mathbf{d} = \varepsilon(\mathbf{e}, \mathbf{h}), \qquad \mathbf{b} = \mu(\mathbf{e}, \mathbf{h}).$$

It is no longer necessary for us to assume any sort of continuity or differentiability on the functions $\varepsilon$ and $\mu$, since our field equations for wave-like solutions are algebraic equations, not differential ones. In particular, they can still include the typical optical media in which there are jump discontinuities at interfaces.

If we then substitute the time+space form of $k$ that is found in (2.5) and the time+space forms of $f$ and $\mathbf{h}$ above in the field equations (2.4) then the latter will split into the form:

(2.12) $$\boxed{\begin{array}{llll} \omega \mathsf{d} = ck_s \wedge \mathbf{h}, & k_s(\mathbf{d}) = 0, & \mathbf{d} = \varepsilon(\mathbf{e},\mathbf{h}), \\ \omega \mathbf{b} = -ck_s \wedge \mathbf{e}, & k_s(\mathbf{b}) = 0, & \mathbf{b} = \mu(\mathbf{e},\mathbf{h}). \end{array}}$$

(We have introduced the 2-form $\mathsf{d} = \#_s \mathbf{d}$ in order to simplify the notation.) We would like to think of this system of algebraic equations as the "algebraic Maxwell equations," although it is important to emphasize that they only pertain to the wave-like solutions to the differential Maxwell equations.

We can already show that the 1-forms $k_s$, $\mathbf{e}$, and $\mathbf{h}$ are linearly independent by taking advantage of the fact that the exterior product of three 1-forms will be non-zero iff they are linearly independent. In particular, from the first equation in the top row above, we will have:

$$k_s \wedge \mathbf{h} \wedge \mathbf{e} = -k_s \wedge \mathbf{e} \wedge \mathbf{h} = \frac{\omega}{c}\mathsf{d} \wedge \mathbf{e} = \frac{\omega}{c}\#_s\mathbf{d} \wedge \mathbf{e} = \frac{\omega}{c} i_\mathbf{d} V_s \wedge \mathbf{e}.$$

However:

$$i_\mathbf{d} V_s \wedge \mathbf{e} = i_\mathbf{d}(V_s \wedge \mathbf{e}) - i_\mathbf{d}\mathbf{e}\, V_s = -\mathbf{e}(\mathbf{d}) V_s,$$

since $V_s \wedge \mathbf{e}$ is a 4-form on a three-dimensional space, and therefore zero. We now have:

(2.13) $$k_s \wedge \mathbf{e} \wedge \mathbf{h} = \frac{\omega}{c} \mathbf{e}(\mathbf{d}) V_s,$$

and since $V_s$ is always non-zero, by definition, the vanishing of the left-hand side will be equivalent to the vanishing of **e** (**d**). However, that basically comes down to the vanishing of the electrostatic energy density in the wave, which is presumed to be non-zero. Thus, the set of three covector fields $\{k_s, \mathbf{e}, \mathbf{h}\}$ are linearly independent and therefore define a coframe.

Note that we could have just as well started from the first equation on the second row of (2.12) and arrived at:

$$(2.14) \qquad k_s \wedge \mathbf{e} \wedge \mathbf{h} = \frac{\omega}{c} \mathbf{h}(\mathbf{b}) V_s,$$

whose vanishing would come down to the vanishing of:

$$(2.15) \qquad \mathbf{h}(\mathbf{b}) = \mathbf{e}(\mathbf{d}),$$

which then defines the magnetostatic energy density of the wave. Note that from (2.7) and (2.10), one will always have:

$$(2.16) \qquad \mathsf{f}(\mathfrak{h}) = 0$$

for any electromagnetic wave. That also implies that its Lagrangian density will vanish identically, as well, although it perhaps best to regard that as more of a constraint than a death blow.

We shall return to a more detailed discussion of the canonical 3-coframes and 3-frames that are defined by any electromagnetic wave, as well as their obvious extensions to 4-coframes and 4-frames in a later section, but for now we shall return to our discussion of the algebraic forms of the equations for electromagnetic waves.

Equations (2.12) can be combined into a purely-electric equation and a purely-magnetic one that serve as essentially "precursors" to the algebraic forms of the wave equations for $E$ and $\mathfrak{H}$. We shall now further restrict the constitutive properties of the medium to have no cross-couplings of electric and magnetic fields, i.e.:

$$(2.17) \qquad \mathbf{d} = \varepsilon(\mathbf{e}), \quad \mathbf{b} = \mu(\mathbf{h}), \quad \text{or} \quad d^i = \varepsilon^{ij} e_j, \quad b^i = \mu^{ij} h_j,$$

In order to obtain the electric equation, one solves the first equation in the bottom row of (2.12) for the bivector field **h** [1]

$$(2.18) \qquad \mathbf{h} = -\frac{c}{\omega} \mu(k_s \wedge \mathbf{e})$$

and substitutes it in the first equation in the top row:

---

[1] The map $\mu$ has been inverted in order for it to take bivectors to 2-forms and 1-forms to vectors.

(2.19)
$$\left(\frac{\omega}{c}\right)^2 \#_s(\mathsf{d}) = -k_s \wedge \#_s[\mu(k_s \wedge \mathsf{e})].$$

Despite its abstruse complexity, that equation is, in fact, linear in the spatial 1-form e, at least, as long as the maps $\varepsilon$ and $\mu$ are linear isomorphisms. In the conventional formulation in the language of vector analysis (e.g., Sommerfeld [2]), the references to $\#_s$ would disappear, and the exterior products would be replaced with vector cross products, i.e.:

(2.20)
$$\left(\frac{\omega}{c}\right)^2 \mathbf{d} = -\mathbf{k} \times \mu(\mathbf{k} \times \mathbf{e}).$$

In order to get the magnetic equation, one needs to find a spatial vector that will correspond to the covector $k_s$. In the conventional formulation, one simply appeals to the unit vector in the direction of the Poynting (co)vector $\mathbf{s} = \mathbf{e} \times \mathbf{h}$. One could express that same vector as $\#_s^{-1}(\mathsf{e} \wedge \mathsf{h})$, but as it turns out, it is more direct to simply define the vector **s** by the properties:

(2.21)
$$k_s(\mathbf{s}) = 1, \qquad \mathsf{e}(\mathbf{s}) = 0, \qquad \mathsf{h}(\mathbf{s}) = 0.$$

One thus avoids the necessity of introducing an auxiliary or background metric in order to define unit vectors. Since $k_s$, e, and h are linearly independent, those equations represent three independent linear equations in three unknowns (e.g., the components of **s** in some choice of frame field). Thus, they can be solved for **s** uniquely. As long as e and h are given beforehand, equations (2.21) also define an invertible linear map from the spatial 1-form $k_s$ to the spatial vector field **s**, which then replaces the need for a Euclidian metric for associating spatial vectors and covectors. We shall return to a detailed discussion of that map in a later section.

We can now convert the algebraic Maxwell equations (2.12), which were expressed in terms of $k_s$, into algebraic equations in terms of **s**. We begin by taking the interior product of the equation for b in (2.12) with **s**, while using the first and second equation in (2.21):

(2.22)
$$\omega\, i_\mathbf{s} \mathsf{b} = -c\, \mathsf{e}.$$

However:
$$i_\mathbf{s} \mathsf{b} = i_\mathbf{s} \#_s \mathbf{b} = i_\mathbf{s} i_\mathbf{b} V_s = \#_s(\mathbf{s} \wedge \mathbf{b}),$$

so we can say that:

(2.23)
$$\frac{c}{\omega} \#_s^{-1} \mathsf{e} = -\mathbf{s} \wedge \mathbf{b}.$$

Taking the interior product of the equation for $\#_s\mathbf{d}$ in (2.12) requires slightly more work since that initially gives:
$$\omega\, i_\mathbf{s} \#_s \mathbf{d} = c\, \#_s \mathbf{h} - c\, k_s\, i_\mathbf{s}(\#_s \mathbf{h}).$$

The left-hand side becomes:
$$\omega \, i_{\mathbf{s}} \, \#_s \mathbf{d} = \omega \, i_{\mathbf{s}} \, i_{\mathbf{d}} \, V_s = \#_s (\omega \, \mathbf{s} \wedge \mathbf{d}) ,$$

while the second term in the right-hand side becomes:
$$- c \, k_s \, i_{\mathbf{s}} \, (\#_s \mathbf{h}) = - c \, k_s \, i_{\mathbf{s}} \, \mathbf{h} = - c \, k_s \, \mathbf{h}(\mathbf{s}) = 0 ,$$

from the properties that defined $\mathbf{s}$. Thus, the equation for $\mathbf{d}$ takes the ultimate form:

(2.24) $$\frac{c}{\omega} \mathbf{h} = \mathbf{s} \wedge \mathbf{d} .$$

Hence, we can summarize the results above in the "dual" algebraic Maxwell equations:

(2.25) $$\boxed{\begin{array}{lll} c\mathbf{e} = -\omega \mathbf{s} \wedge \mathbf{b}, & \mathbf{e}(\mathbf{s}) = 0, & \mathbf{d} = \varepsilon(\mathbf{e}), \\ c\mathbf{h} = \omega \mathbf{s} \wedge \mathbf{d}, & \mathbf{h}(\mathbf{s}) = 0, & \mathbf{b} = \mu(\mathbf{h}). \end{array}}$$

(Once again, we have introduced the bivector field $\mathbf{e}$, for clarity.) The form of these equations should be compared to that of (2.12).

We can apply an argument that is analogous to the one that was used to show that the triple of 1-forms $\{k_s, \mathbf{e}, \mathbf{h}\}$ is linearly independent, and therefore defines a 3-coframe field, to show that the triple of vector fields $\{\mathbf{s}, \mathbf{d}, \mathbf{b}\}$ is also linearly independent, and therefore defines a 3-frame field. Namely, one simply shows that the 3-vector $\mathbf{s} \wedge \mathbf{d} \wedge \mathbf{b}$ never vanishes.

An analogous process to the one above can be applied to those equations. Namely, one solves the first equation in the second row for $\mathbf{b} = \mu(\mathbf{h})$ and substitutes that in the first equation in the first row, which will then give:

(2.26) $$\boxed{\left(\frac{c}{\omega}\right)^2 \#_s^{-1}(\mathbf{e}) = -\mathbf{s} \wedge \mu^{-1}[\#_s(\mathbf{s} \wedge \mathbf{d})] .}$$

This can then be compared with the dual equation (2.19) for $k_s$. Basically, everything has been inverted since one can convert one equation into its dual by the following replacements:

(2.27) $$\frac{\omega}{c} \leftrightarrow \frac{c}{\omega}, \qquad \#_s \leftrightarrow \#_s^{-1}, \qquad \mathbf{e} \leftrightarrow \mathbf{d}, \qquad k_s \leftrightarrow \mathbf{s}, \qquad \mu^{-1} \leftrightarrow \mu .$$

Indeed, the way that the relationship between $k_s$ and $\mathbf{s}$ amounts to an inversion can be seen from the defining relationship $k_s(\mathbf{s}) = 1$. If one chooses a coframe in which the components of $k_s$ are $(k, 0, 0)$ then a possible choice of corresponding components for $\mathbf{s}$ might be $(1/k, 0, 0)$.

**3. Fresnel's surfaces.** – For a typical optical medium, since the magnetic properties are assumed to linear, isotropic, and homogeneous, the tensor $\mu$ will take the form of a conformal transformation of the Euclidian metric on space, so it will take the form of $\mu_0 \delta^{ij}$. One can then use that fact in order to define the principal velocities of propagation for electromagnetic waves and the principal indices of refraction for the medium itself can be defined by the expressions:

$$(3.1) \qquad (c^i)^2 = \frac{1}{\varepsilon_i \mu_0}, \qquad n_i^2 = \frac{\varepsilon_i \mu_0}{\varepsilon_0 \mu_0} = \frac{\varepsilon_i}{\varepsilon_0} = \left(\frac{c}{c^i}\right)^2,$$

in which the $\varepsilon_i$ are the principal values of $\varepsilon^{ij}$, i.e., the diagonal elements in a principal frame.

They, in turn, suggest that one can define metrics on the cotangent and tangent spaces that have a more kinematical character than an electromagnetic one by way of:

$$(3.2) \qquad (c^{ij})^2 = \frac{1}{\mu_0} \varepsilon^{ij}, \qquad n_{ij}^2 = \frac{1}{\varepsilon_0} \varepsilon_{ij}.$$

With those alterations, the characteristic equation (2.19) will take the form:

$$(3.3) \qquad \frac{1}{\mu_0}\left(\frac{\omega}{c}\right)^2 \#_s(\mathbf{d}) = - k_s \wedge \#_s [\delta^{-1}(k_s \wedge \mathbf{e})],$$

in which the symbol $\delta^{-1}$ refers to the linear isomorphism of 2-forms with bivectors that is defined by the Euclidian metric.

In order to calculate the actual characteristic polynomial that this equation implies, it helps to revert to the vector analytic form of it that one might find in, say, Sommerfeld [**2**]:

$$(3.4) \qquad \frac{1}{\mu_0}\left(\frac{\omega}{c}\right)^2 \mathbf{d} = - \mathbf{k} \times (\mathbf{k} \times \mathbf{e}),$$

and with a well-known identity for the double cross product, that will become:

$$(3.5) \qquad \frac{1}{\mu_0}\left(\frac{\omega}{c}\right)^2 \mathbf{d} = - (\mathbf{k} \cdot \mathbf{e})\, \mathbf{k} + (\mathbf{k} \cdot \mathbf{k})\, \mathbf{e}.$$

When everything is replaced with its components with respect to a natural frame field on space, and some rearrangements are made, that will become the matrix equation:

$$(3.6) \qquad 0 = [(\omega/c)^2 \varepsilon^{ij}/\mu_0 + \delta^{im}\delta^{jn} k_m k_n - k^2 \delta^{ij}] e_j.$$

The $1/\mu_0$ can be combined with $\varepsilon^{ij}$ to give $(c^2)^{ij}$, which will take the form of a diagonal matrix $\text{diag}[c_1^2, c_2^2, c_3^2]$ in a principal frame for $\varepsilon^{ij}$. The diagonal entries are the squares of what one calls the *principal velocities* for the propagation of electromagnetic waves. If one also introduces the notation $k_0 = \omega/c$, to simplify, then the last equation will take the form:

$$(3.7) \qquad 0 = [k_0^2 (c^2)^{ij} + \delta^{im} \delta^{jn} k_m k_n - k^2 \delta^{ij}] e_j .$$

In order for non-trivial electromagnetic waves to exist, that equation must admit non-zero solutions for $e_j$. Thus, the matrix in the square brackets cannot be invertible, so its determinant must vanish. That gives the characteristic equation for the frequency-wave number 2-form $k$, whose components are $(k_0, k_i)$:

$$(3.8) \qquad 0 = \det[k_0^2 (c^2)^{ij} + \delta^{im} \delta^{jn} k_m k_n - k^2 \delta^{ij}] \equiv \mathcal{D}(k_0, k_i) .$$

When expanded out in full, we will get the homogeneous quartic polynomial in $k_0$, $k_i$):

$$(3.9) \qquad \mathcal{D}(k_0, k_i) = k_0^4 - B(k_i) k_0^2 + C(k_i),$$

in which we have defined the coefficients:

$$(3.10) \qquad \begin{cases} B(k_i) = (c_3^2 + c_2^2) k_1^2 + (c_1^2 + c_3^2) k_2^2 + (c_1^2 + c_2^2) k_3^2, \\ C(k_i) = (c_2^2 c_3^2 k_1^2 + c_3^2 c_1^2 k_2^2 + c_1^2 c_2^2 k_3^2) k^2. \end{cases}$$

Although the matrix in question is three-by-three, which would suggest that the polynomial should be sextic, the fact that electromagnetic waves are presumed to be transverse leads to a factoring of that sextic into a quadratic polynomial that is always equal to a constant times the quartic polynomial above.

The dispersion law for electromagnetic waves in this particular medium will then take the form:

$$(3.11) \qquad \mathcal{D}(k_0, k_i) = 0,$$

which defines a quartic hypersurface in each cotangent space to space-time.

Since the polynomial $\mathcal{D}(k_0, k_i)$ takes the form of a quadratic equation for $k_0^2$, it can be solved for $k_0^2$ (although not uniquely, in general):

$$(3.12) \qquad k_0^2 = B(k_i) \pm \sqrt{B^2(k_i) - 4C(k_i)} .$$

This has the effect that whenever one chooses a wave number $k_i$, one will generally get two distinct frequencies $k_0$ that correspond to it. Thus, when one divides $k$ by $k_0$, one will get two distinct indices of refraction:

$$(3.13) \qquad n = \frac{k}{k_0} = \frac{kc}{\omega} = \frac{c}{v}$$

for the direction of $k_i$. The $v$ in that expression refers to the phase velocity of the propagation of the tangent plane to the spatial wave surface that is defined by the Fresnel wave surface. The existence of a double root for $k_0^2$ when $k_i$ is specified is referred to as *birefringence* or *double refraction*.

It is possible for the quartic polynomial $\mathcal{D}(k_0, k_i)$ to factor into a product of quadratic polynomials of Minkowski type, which is then a situation that is often referred to as *bimetricity* [**8**]. The condition for that factorization to be possible is that the discriminant of the dispersion polynomial, namely:

$$(3.14) \quad B^2(k_i) - 4C(k_i)$$
$$= [(c_3^2 + c_2^2)k_1^2 + (c_1^2 + c_3^2)k_2^2 + (c_1^2 + c_2^2)k_3^2]^2 - 4(c_2^2 c_3^2 k_1^2 + c_3^2 c_1^2 k_2^2 + c_1^2 c_2^2 k_3^2)k^2,$$

must be positive.

That situation would be typical for *uniaxial* optical media, for which two of the principal dielectric constants (hence, two of the principal velocities) are equal, while the third one is unequal. For instance, if $c_1 = c_2 \neq c_3$ then:

$$(3.15) \qquad B(k_i) = (c_1^2 + c_3^2)(k_1^2 + k_2^2) + 2c_1^2 k_3^2, \quad C(k_i) = [c_1^2 c_3^2 (k_1^2 + k_2^2) + c_1^4 k_3^2]k^2,$$

and the discriminant would become:

$$(3.16) \qquad B^2 - 4C = (c_1^2 - c_3^2)^2 (k_1^2 + k_2^2)^2,$$

which is clearly positive in all cases. The roots of the dispersion polynomial (i.e., the two values of $k_0^2$) would then be:

$$(3.17) \qquad k_+^2 = 2c_1^2 k^2, \qquad k_-^2 = 2c_3^2 (k_1^2 + k_2^2) + c_1^2 k_3^2,$$

the first of which defines a sphere in $k$-space when $k_+$ is constant, while the second one defines an ellipsoid when $k_-$ is held constant.

The polynomial $\mathcal{D}(k_0, k_i)$ then factors into the product:

$$(3.18) \qquad \mathcal{D}(k_0, k_i) = (k_0^2 - k_+^2)(k_0^2 - k_-^2).$$

The dispersion polynomial will further degenerate to the square of a Minkowski quadratic when the discriminant vanishes identically, so there is a double root for $k_0^2$. That happens in the isotropic case, where $c_1 = c_2 = c_3 = c$, so the coefficients reduce to $B(k_i) = 2c^2(k_1^2 + k_2^2 + k_3^2) = 2c^2 k^2$ and $C(k_i) = c^4 k^4$, and the quartic polynomial takes the form:

$$\mathcal{D}(k_0, k_i) = (k_0^2 - k^2)^2, \tag{3.19}$$

which is, in fact, the square of the quadratic polynomial that gives one the light cones of Minkowski space.

In general, the vanishing of the discriminant will only define certain directions in which the roots coalesce, which are then associated with the optical phenomenon of *conical refraction*, since the tangent vector spaces degenerate to tangent cones at those singular points of the surface.

In order to get from the space-time dispersion law to the spatial wave surface, one treats the components $(k_0, k_i)$ as the homogeneous coordinates of a point in the projective space $\mathbb{RP}^3$, and makes them to their corresponding inhomogeneous coordinates:

$$N_i = \frac{k_i}{k_0} = \frac{k_i c}{\omega} = \frac{c}{v^i}. \tag{3.20}$$

Thus, the inhomogeneous coordinates $N_i$ take the form of indices of refraction, although not generally the principal indices that are associated with $\varepsilon^{ij}$, since $k_i$ can point in any spatial direction.

When one divides the dispersion polynomial $\mathcal{D}(k_0, k_i)$ by $k_0^4$, it will give a spatial quartic polynomial for the inhomogeneous coordinates $N_i$:

$$\mathcal{D}'(N_i) = 1 - B(N_i) + C(N_i) \tag{3.21}$$
$$= 1 - (c_3^2 + c_2^2) N_1^2 - (c_1^2 + c_3^2) N_2^2 - (c_1^2 + c_2^2) N_3^2 + (c_2^2 c_3^2 N_1^2 + c_3^2 c_1^2 N_2^2 + c_1^2 c_2^2 N_3^2) N^2,$$

which can be expressed more concisely in the traditional form:

$$0 = \frac{N_1^2}{v^2 - c_1^2} + \frac{N_2^2}{v^2 - c_2^2} + \frac{N_3^2}{v^2 - c_3^2} \tag{3.22}$$

when one replaces $N$ with $c/v$.

Note that it is irrelevant whether the covector $N_i$ has been normalized to a unit vector since the right-hand side of the last equation is homogeneous in its components. Hence, any covector that defines the same direction in the cotangent space at a point in question will still be associated with the same value of $v$. Thus, we are still justified in pursuing our pre-metric agenda.

The dual equation (2.26) can be treated analogously, and one will get a characteristic equation for the space-time tangent vectors $(s^0, s^i)$, in which one defines $s^0 = c/\omega$. In order to circumvent the repetition of the previous sequence of calculations, one can use the aforementioned inversion transformation (2.27) to get a corresponding quartic polynomial in the tangent spaces:

$$(3.23) \qquad \mathfrak{D}(s^0, s^i) = s_0^4 - \mathfrak{B}(s^i) s_0^2 + \mathfrak{C}(s^i),$$

in which:

$$(3.24) \qquad s^0 \equiv \frac{1}{k_0} = \frac{c}{\omega},$$

and

$$(3.25) \qquad \begin{cases} \mathfrak{B}(s^i) = (n_3^2 + n_2^2)(s^1)^2 + (n_1^2 + n_3^2)(s^2)^2 + (n_1^2 + n_2^2)(s^3)^2, \\ \mathfrak{C}(s^i) = [n_2^2 n_3^2 (s^1)^2 + n_3^2 n_1^2 (s^2)^2 + n_1^2 n_2^2 (s^2)^2] s^2. \end{cases}$$

One can then project from the homogeneous coordinates $(s^0, s^i)$ to the inhomogeneous coordinates:

$$(3.26) \qquad S^i = \frac{s^i}{s^0},$$

and ultimately one will get the traditional form of the *Fresnel ray surface*:

$$(3.27) \qquad 0 = \frac{(S^1)^2}{n^2 - n_1^2} + \frac{(S^2)^2}{n^2 - n_2^2} + \frac{(S^3)^2}{n^2 - n_3^2}.$$

Once again, since the right-hand side is homogeneous in the components $S^i$, it is irrelevant whether that vector has been normalized to a unit vector.

Whereas the Fresnel wave (or normal) surface associated one or two phase velocities with a given covector in the form of the normal covector to the tangent planes to the wave surface, the ray surface associates one or two *group* velocities (by way of *n*) with a given tangent vector, which is then assumed to be tangent to a light ray through the point that the vector is tangent to.

Although it is conventional to say that the direction of the light ray through a given point *is not* collinear with the direction of the normal to the wave front through that same point, we now see that in the absence of a metric that would serve as an independent map from tangent spaces to cotangent spaces, one cannot actually compare their directions. That goes back to the fact that we are also treating **D** as a tangent vector, while *E* is a covector, so it is also absurd to define the angle between them, as is customary. Although we do have two linear isomorphisms of the tangent spaces and cotangent spaces that are defined by the dielectric strength tensor and the magnetic permeability tensor, the former would map *E* to **D** identically, so that angle would always vanish. However, one can still deal with the scalar function that is obtained from *E* (**D**), which relates to the electrostatic energy density, but in the absence of a metric so that we could normalize $E$ and **D**, one cannot speak of that number as defining the cosine of an angle.

**4. Canonical frames and spatial metrics defined by electromagnetic waves.** – As we have seen, the very nature of wave-like solutions to the Maxwell equations defines a canonical linear coframe on spatial cotangent spaces in the form of ($k_s$, e, h) and a corresponding frame on spatial tangent spaces that is defined by (**s**, **d**, **b**). From the properties of the medium and the algebraic Maxwell equations, along with the definition of **s**, one can then compile the following matrix of values for the evaluation of the 1-forms on the tangent vectors:

$$k_s(\mathbf{s}) = 1, \quad k_s(\mathbf{d}) = 0, \qquad k_s(\mathbf{b}) = 0,$$
$$\mathsf{e}(\mathbf{s}) = 0, \quad \mathsf{e}(\mathbf{d}) = \varepsilon(\mathsf{e},\mathsf{e}), \quad \mathsf{e}(\mathbf{b}) = 0,$$
$$\mathsf{h}(\mathbf{s}) = 0, \quad \mathsf{h}(\mathbf{d}) = 0, \qquad \mathsf{h}(\mathbf{b}) = \mu(\mathsf{h},\mathsf{h}).$$

In order to prove the statements that e (**b**) and h (**d**) vanish, one needs only to refer to equations (2.25). One then finds that:

$$\mathsf{e}\,(\mathbf{b}) = i_\mathbf{b}\,\mathsf{e} = i_\mathbf{b}\, \#_s\mathsf{e} = i_\mathbf{b}\, i_\mathsf{e}\, V_s = -\#_s(\mathsf{e} \wedge \mathbf{b}) = 0,$$
$$\mathsf{h}\,(\mathbf{d}) = i_\mathbf{d}\,\mathsf{h} = i_\mathbf{d}\, \#_s\mathsf{h} = i_\mathbf{d}\, i_\mathsf{h}\, V_s = -\#_s(\mathbf{d} \wedge \mathsf{h}) = 0.$$

It is the last step in each row that follows from the first equations in each row of (2.25).

Whenever one is given a linear frame in a vector space $V$, one can always use that frame to define a metric on that space by simply *defining* the given frame to be orthonormal with some chosen signature type. For instance, if one has a coframe $\{\theta^i, i = 1, \ldots, n\}$ on the dual space $V^*$ then one can define a Riemannian metric $g$ on $V$ by way of:

(4.1) $$g = \delta_{ij}\,\theta^i\,\theta^j.$$

One could choose any other signature type, such as the Minkowski type $\{+, -, -, -\}$, and defined a pseudo-Riemannian metric the same way. Dually, if one is given a frame $\{e_i, i = 1, \ldots, n\}$ on $V$ then one can define a Riemannian metric on $V^*$ by way of:

(4.2) $$g = \delta^{ij}\,\mathbf{e}_i\,\mathbf{e}_j.$$

Since we have a coframe $\{k_s, \mathsf{e}, \mathsf{h}\}$ on the spatial cotangent spaces and a frame $\{\mathbf{s}, \mathbf{d}, \mathbf{b}\}$ on the spatial tangent spaces, we can use the former to define a metric on the spatial tangent spaces and the latter to define a metric on the cotangent spaces. The obvious question to ask is whether one is essentially the inverse of the other, as would be the usual case. That, in turn, comes down to the question of whether the frame and coframe are reciprocal, and we shall see shortly, they can both be "normalized" in a way that will make that true.

The fact that the second-rank covariant tensor fields $\varepsilon$ and $\mu$ are symmetric and non-degenerate means that they define metrics on the tangent spaces, and by inversion, metrics on the cotangent spaces. Since they are presumed to be positive-definite, from the typical constitutive properties of electromagnetic material, they also define Riemannian metrics. However, they are not generally isometric, since they can, in fact, differ in terms of their isotropy properties with respect to a given frame. That is, a principal frame for one might not be a principal frame for the other, and even if it is, the principal values can differ.

However, one can use the two metrics in order to define norms on the various electric and magnetic vectors and covectors:

$$(4.3) \qquad \|\mathbf{e}\|_\varepsilon^2 = \varepsilon(\mathbf{e}, \mathbf{e}), \qquad \|\mathbf{d}\|_\varepsilon^2 = \varepsilon^{-1}(\mathbf{d},\mathbf{d}),$$

$$(4.4) \qquad \|\mathbf{h}\|_\mu^2 = \mu(\mathbf{h}, \mathbf{h}), \qquad \|\mathbf{b}\|_\mu^2 = \mu^{-1}(\mathbf{b},\mathbf{b}).$$

Note that both of those spatial metrics have immediate physical interpretations: The electric metric (or really, its norm-squared) gives something proportional to the energy density of the electric field, while the magnetic metric gives something proportional to the energy density of the magnetic field. Thus, they are not mere auxiliary mathematical constructions in the eyes of physics.

Since:

$$(4.5) \qquad \mathbf{e}(\mathbf{d}) = \varepsilon(\mathbf{e}, \mathbf{e}) = \varepsilon^{-1}(\mathbf{d},\mathbf{d}) \quad \text{and} \quad \mathbf{h}(\mathbf{d}) = \mu(\mathbf{h}, \mathbf{h}) = \mu^{-1}(\mathbf{b},\mathbf{b}),$$

one will have:

$$(4.6) \qquad \|\mathbf{e}\|_\varepsilon = \|\mathbf{d}\|_\varepsilon, \qquad \|\mathbf{h}\|_\mu = \|\mathbf{b}\|_\mu.$$

As a result, one can define unit vectors and covectors in the directions of the electric and magnetic vectors and covectors:

$$(4.7) \qquad \hat{\mathbf{e}} = \frac{\mathbf{e}}{\|\mathbf{e}\|_\varepsilon}, \quad \hat{\mathbf{d}} = \frac{\mathbf{d}}{\|\mathbf{d}\|_\varepsilon}, \quad \hat{\mathbf{h}} = \frac{\mathbf{h}}{\|\mathbf{h}\|_\mu}, \quad \hat{\mathbf{b}} = \frac{\mathbf{b}}{\|\mathbf{b}\|_\mu}.$$

With those normalizations the matrix above will now take the form:

$$k_s(\mathbf{s}) = 1, \quad k_s(\hat{\mathbf{d}}) = 0, \quad k_s(\hat{\mathbf{b}}) = 0,$$
$$\hat{\mathbf{e}}(\mathbf{s}) = 0, \quad \hat{\mathbf{e}}(\hat{\mathbf{d}}) = 1, \quad \hat{\mathbf{e}}(\hat{\mathbf{b}}) = 0,$$
$$\hat{\mathbf{h}}(\mathbf{s}) = 0, \quad \hat{\mathbf{h}}(\hat{\mathbf{d}}) = 0, \quad \hat{\mathbf{h}}(\hat{\mathbf{b}}) = 1.$$

Thus, the frame $(\mathbf{s}, \hat{\mathbf{d}}, \hat{\mathbf{b}})$ and the coframe $(k_s, \hat{\mathbf{e}}, \hat{\mathbf{h}})$ are reciprocal to each other. As a result, the association of the frame $(\mathbf{s}, \hat{\mathbf{d}}, \hat{\mathbf{b}})$ in any spatial tangent space $T_x\Sigma$ with the coframe $(k_s, \hat{\mathbf{e}}, \hat{\mathbf{h}})$ in its dual cotangent space $T_x^*\Sigma$ will define a linear isomorphism between the two vector spaces. If the components of a spatial tangent vector **v** with respect to the frame (**s**, **d̂**, **b̂**) are $(v^1, v^2, v^3)$ then the components of its dual covector with respect to the coframe $(k_s, \hat{e}, \hat{h})$ will also be $(v^1, v^2, v^3)$, so the map that takes one set of components to the other is essentially that of transposing a column matrix to a row matrix. One can also define that linear isomorphism to be one that is associated with a Euclidian metric on the spatial tangent space $T_x\Sigma$ by defining the frame (**s**, **d̂**, **b̂**) to be orthonormal. Thus, the components of the metric with respect to that frame will be $\delta_{ij}$.

Note that by now we have defined three spatial metrics of an electromagnetic nature: The electric one $\varepsilon$, the magnetic one $\mu$, and the electromagnetic one that was just defined by the wave itself. However, there is an essential difference between them, since the first two are properties of the medium itself, regardless of whether an electromagnetic wave is present, while the last one exists only at points that have been excited by the passage of an electromagnetic wave.

One can express the reciprocal frame and coframe field that have thus been defined in matrix form. Namely, the frame in the tangent spaces can be represented by a row vector whose elements are column vectors [**s** | **d** | **b**], while the reciprocal coframe in the cotangent spaces takes the form of the column vector whose elements are row vectors $[k_s | e | h]^T$. Their reciprocal relationship that defined above can then be expressed as the matrix equation:

$$(4.8) \quad \begin{bmatrix} k_s \\ \hline e \\ \hline h \end{bmatrix} [\mathbf{s} \mid \mathbf{d} \mid \mathbf{b}] = \begin{bmatrix} 1 & 0 & 0 \\ 0 & \|e\|_\varepsilon^2 & 0 \\ 0 & 0 & \|h\|_\mu^2 \end{bmatrix}.$$

The vectors involved can also be normalized using the metrics defined by $\varepsilon$ and $\mu$ to make the right-hand side of that take the form of an identity matrix.

Since we have assumed a time+space decomposition of the tangent and cotangent bundles in order to define the spatial metrics, it is tempting to extend the spatial metrics to spacetime metrics in some way. However, if we recall that the true origin of the Lorentzian (i.e., Minkowski space) metric was in the dispersion law for the propagation of electromagnetic waves in the classical vacuum then we will see that since we also started with a more elaborate dispersion law than the one that gives Minkowski space as only a degenerate case, it would be physically meaningless to make any such extension of the spatial metrics, as opposed to simply returning the quartic surfaces that we started with.

Nonetheless, there is some utility to extending the spatial frame (**s**, **d**, **b**) and the spatial coframe ($k_s$, e, h) to space-time frames since that would define a linear isomorphism of the four-dimensional tangent spaces with the corresponding four-dimensional cotangent spaces. One simply recalls that the time+space decomposition was effected by a pair (**t**, $\tau$) that consists of a non-zero vector field **t** and a non-zero covector field $\tau$ that are transverse, so not only does one have $\tau(\mathbf{t}) \neq 0$, but also that **t** is transverse to the spatial hyperplanes $\Sigma$ (*M*) that are annihilated by $\tau$, and $\tau$ is transverse to the spatial hyperplanes $\Sigma^*$ (*M*) that are annihilated by **t**. Thus, the four-frame (**t**, **s**, **d**, **b**) will span the four-dimensional tangent spaces in $T(M)$, while the four-coframe ($\tau$, $k_s$, **e**, **h**) will span the four-dimensional cotangent spaces in $T^*(M)$.

**5. Discussion.** – It now becomes clearer that the dual role that is played by the Lorentzian metric on the spacetime manifold, namely, that in addition to being an artifact of the dispersion law for electromagnetic waves in the classical electromagnetic vacuum, it also defines a linear isomorphism of each tangent space with its corresponding cotangent space, breaks down into separate problems when one goes to more general (i.e., quartic) dispersion laws. That is, the linear isomorphism comes about only as a result of the fact that electromagnetic waves define canonical coframes in the cotangent spaces and their reciprocal frames in the tangent spaces.

There are various directions to pursue on the name of generalizing the geometry that is implied by electromagnetic waves that are defined by generalizing the restrictions that were imposed upon the constitutive law above. For instance, one could examine the effect of including the cross-couplings that come from the Faraday effect, the Kerr effect, and optical activity. Ultimately, since one expects that the field strengths in the close neighborhood of elementary charges should be high enough that nonlinear constitutive laws would need to be considered, in particular, some sort of quasi-classical model for vacuum polarization, the extension to nonlinear media would seem unavoidable. One might even speculate upon whether the nonlocality of quantum entanglement is somehow related to the nonlocality of the constitutive law, i.e., does it involve dispersion in the other sense of the word?

Since two of the covector fields in the canonical coframe of an electromagnetic wave spanned the tangent spaces to the isophase surfaces, one might wonder how that construction might be generalized to other types of waves, such as mechanical or gravitational, or even matter waves.

Eventually, one must try to include polarization in the geometry, in the sense that relates to the plane of polarization. Typically, that involves the introduction of complex electromagnetic fields, but one might consider that complex numbers are simply a convenient way of representing planar rotations, and that their exist real forms of the equations, such as the discussion in Kline and Kay [9].

———

**Appendix**

Here are the details of the calculation of the divergence of $\mathfrak{H}$:

$$\mathrm{div}\, \mathfrak{H} = \frac{1}{c} \mathrm{div}\,(\partial_t \wedge \mathbf{D}) + \mathrm{div}\,(\#_s^{-1} H)\ .$$

First of all:

$$\mathrm{div}\,(\partial_t \wedge \mathbf{D}) = \#^{-1}\, d_\wedge \# (\partial_t \wedge \mathbf{D}) = c\, \#^{-1}\, d_\wedge \#_s\, \mathbf{D} = c\, \#^{-1}\, (dt \wedge \partial_t \#_s\, \mathbf{D} + d_{\wedge s}\, \#_s\, \mathbf{D})$$

Now:

$$\#^{-1}(dt \wedge \partial_t \#_s \mathbf{D}) = i_{dt \wedge \partial_t \#_s \mathbf{D}} \mathbf{V} = i_{dt \wedge \partial_t \#_s \mathbf{D}} \frac{1}{c} \partial_t \wedge \mathbf{V}_s = \frac{1}{c} i_{\partial_t \#_s \mathbf{D}} \mathbf{V}_s = \frac{1}{c} i_{\#_s \partial_t \mathbf{D}} \mathbf{V}_s = \frac{1}{c} \#_s^{-1} \#_s \partial_t \mathbf{D}$$

$$= \frac{1}{c} \partial_t \mathbf{D},$$

in which we have defined:

$$\mathbf{V}_s = \partial_1 \wedge \partial_2 \wedge \partial_3 = \frac{1}{3!} \varepsilon^{ijk} \partial_i \wedge \partial_j \wedge \partial_k.$$

(Note that we have commuted the order of $\partial_t$ and $\#_s$ since $\mathbf{V}_s$ is independent of time.) Therefore:

$$c\,\#^{-1}(dt \wedge \partial_t \#_s \mathbf{D}) = \partial_t \mathbf{D}.$$

We next address:

$$c\,\#^{-1}(d_{\wedge s} \#_s \mathbf{D}) = c\,i_{d_{\wedge s} \#_s \mathbf{D}} \mathbf{V} = c\,i_{d_{\wedge s} \#_s \mathbf{D}} \frac{1}{c} \partial_t \wedge \mathbf{V}_s = -(i_{d_{\wedge s} \#_s \mathbf{D}} \mathbf{V}_s)\partial_t = -(\#_s^{-1} d_{\wedge s} \#_s \mathbf{D})\partial_t$$

$$= -\operatorname{div}_s \mathbf{D}\,\partial_t,$$

i.e.:

$$c\,\#^{-1}(d_{\wedge s} \#_s \mathbf{D}) = -\operatorname{div}_s \mathbf{D}\,\partial_t.$$

That makes:

(A.1)     $\operatorname{div}(\partial_t \wedge \mathbf{D}) = -\operatorname{div}_s \mathbf{D}\,\partial_t + \partial_t \mathbf{D}.$

We now turn to:

$$\operatorname{div}(\#_s^{-1} H) = \#^{-1} d_\wedge \# \cdot \#_s^{-1} H = \#^{-1}[dt \wedge \partial_t (\# \cdot \#_s^{-1} H) + d_{\wedge s}(\# \cdot \#_s^{-1} H)].$$

We need to evaluate the expression $\# \cdot \#_s^{-1} H$. Now:

$$\#_s^{-1} H = i_H \mathbf{V}_s$$

is a purely-spatial bivector field; call it $\mathbf{b}$. Thus:

$$\#\mathbf{b} = i_\mathbf{b} V = i_\mathbf{b}(c\,dt \wedge V_s) = -c\,i_\mathbf{b}(V_s \wedge dt) = -c\,\#_s \mathbf{b} \wedge dt = -c\,\#_s \#_s^{-1} H \wedge dt = -c\,H \wedge dt,$$

so

$$\# \cdot \#_s^{-1} H = c\,dt \wedge H.$$

Thus:

$$\operatorname{div}(\#_s^{-1} H) = c\,\#^{-1}[dt \wedge dt \wedge \partial_t H + d_{\wedge s}(dt \wedge H)] = -c\,\#^{-1}[dt \wedge d_{\wedge s} H] = -c\,i_{dt \wedge d_{\wedge s} H} \mathbf{V}$$

$$= -c\, i_{dt \wedge d_{\wedge s} H}\, \frac{1}{c} \partial_t \wedge \mathbf{V}_s = i_{d_{\wedge s} H}\, i_{dt}\, (\partial_t \wedge \mathbf{V}_s) = i_{d_{\wedge s} H}\, \mathbf{V}_s = \#_s^{-1}\, d_{\wedge s} H,$$

or

(A.2) $$\operatorname{div}(\#_s^{-1} H) = \#_s^{-1}\, d_{\wedge s} H.$$

Thus, upon combining (A.1) and (A.2), we will ultimately get:

$$\operatorname{div} \mathfrak{H} = -\frac{1}{c} \partial_t \mathbf{D} + \#_s^{-1}(d_{\wedge s} H) + \operatorname{div}_s \mathbf{D}\, \frac{1}{c} \partial_t.$$

____

## References

1. D. H. Delphenich, "Line geometry and electromagnetism, I-IV," arXiv:1309.2933, 1311.6766, 1404.4330, 1610.06822.
2. A. Sommerfeld, *Optics*, Lectures on Theoretical Physics, v. IV, Academic Press, New York, 1954.
3. L. D. Landau, E. M. Lifschitz, and L. P. Pitaevskii, *Electrodynamics of Continuous Media,* 2nd ed., Pergamon, Oxford, 1984.
4. Geometrical and topological methods in mathematical physics:
    W. Thirring, *Classical Field Theory,* Springer, Berlin, 1978.
    T. Frenkel, *The Geometry of Physics: an introduction*, Cambridge University Press, Cambridge, 1997.
5. Optics:
    P. Drude, *The Theory of Optics*, Dover, Mineola, NY, 1959; English translation of *Lehrbuch der Optik* (Leipzig, 1900) by C. R. Mann and R. A. Millikan.
    M. Born and E. Wolf, *Principles of Optics,* Pergamon, Oxford, 1980.
6. Pre-metric electromagnetism:
    F. W. Hehl and Y. Obukhov, *Foundations of Classical Electrodynamics,* Birkhäuser, Boston, 2003.
    D. H. Delphenich, *Pre-metric Electromagnetism*, Neo-classical Press, 2009.
7. Time+space splitting:
    D. H. Delphenich, "Nonlinear connections and 1+3 splittings of spacetime," arXiv:gr-qc/0702115.
    D. H. Delphenich, "Transverse geometry and physical observers," arXiv:0711.2033.
    D. H. Delphenich, "On the local integrability of almost-product structures defined by space-time metrics," arXiv:1607.03839.
8. M. Visser, C. Barcelo, and S. Liberati, "Bi-refringence versus bi-metricity," arXiv: gr-qc/0204017.
9. M. Kline and I. W. Kay, *Electromagnetic Theory and Geometrical Optics*, Wiley-Interscience, New York, 1965.

__________